# Towards a Reference Model for Open Access and Knowledge Sharing, Lessons from Systems Research

Paola Di Maio
ISTCS.org
Edinburgh

**Abstract**
The Open Access Movement has been striving to grant universal unrestricted access to the knowledge and data outputs of publicly funded research. leveraging the real time, virtually cost free publishing opportunities offered by the internet and the web. However, evidence suggests that in the systems engineering domain open access policies are not widely adopted. This paper presents the rationale, methodology and results of an evidence based inquiry that investigates the dichotomy between policy and practice in Open Access (OA) of systems engineering research in the UK, explores entangled dimensions of the problem space from a socio-technical perspective, and issues a set of recommendations, including a reference model outline for knowledge sharing in systems research.

## 1. Introduction

The web provides without doubt the most efficient mechanism to exchange explicit knowledge, as long as this is codified and represented using appropriate formalisms and supporting artifacts. A wealth of research, platforms and technologies has been produced in recent decades much of it thanks to considerable public investment, yet despite the availability of good practices and no shortage of openly available knowledge sharing tools and platforms, much knowledge produced with taxpayer's money is still not shared, or only notionally shared, and there is no indication that the uptake of Open Access policies is actually monitored. The research aims to:

- identify policies and practices that regulate the explicit sharing of knowledge generated by publicly funded research in the UK, the body specifically in relation to systems engineering research,
- evaluate to what extent, and via which mechanisms and behaviors, the adoption of OA policies and knowledge sharing artifacts and processes are adopted, with specific focus for this study is systems engineering research in the UK

- devise examples of explicit knowledge models and artifacts to facilitate he codification and sharing of systems knowledge.

## 2. Contribution and Paper Organisation

This paper aims to identify and address a possible gap between the and the practice in Open Access in Scholarly research. It is organized as follows:

**Knowledge sharing challenges:** introductory discussion, and scope of the work.
**Evidence and what works:** overall research approach and Evidence Based Research, and an outline of the research plan.
**Knowledge sharing behaviours and NECTISE:** segmentation of the research field, and a case study, and ethnographic observations
**Open access and knowledge sharing:** filling the gap between two research strands
**The surveying instruments:** introducing Open Access Monitor and the Knowledge Audit Framework.
**Preliminary findings:** the initial results of this research to date.

## 3. Knowledge Sharing Challenges

Knowledge is one of the most valuable resources for individuals and organizations. Scholarly discussion on 'What is knowledge' (as opposed to information for example) are ongoing. For the purpose of this research the 'data-information-knowledge' classical distinction proposed in different versions by various authors is accepted (Ackoff; Bellenger; Sveiby; Davenport and Prusak). 'Knowledge sharing' is intended as making knowledge resources available on the web for free and unrestricted access, use and reuse. Despite decades of research and practice in knowledge management, knowledge sharing and reuse remains elusive, fragmented and compartmentalized (Mandl, et al). This in our hypothesis is due to systemic causes, which we address in our proposed approach. Several disciplines have been converging in recent years to facilitate and increase knowledge exchanges. Pervasive web based technologies have removed many of the physical barriers to knowledge sharing. However many challenges still inhibit optimal knowledge flows. This research targets the challenges associated with accessing knowledge that has been generated using public funding via public research councils in the UK: the UK is one of the countries perceived to be leading the 'freedom of information good practice' and which has been spearheading 'open access' since the early days, yet according to evidence gathered in our research, there are still many gaps in the practical uptake. In particular, since this research originated in the NECTISE research framework (Networked Enabled Capabilities for Systems Engineering) the current scope of the inquiry is primarily on systems engineering research in the UK, therefore constraining the focus of the analysis mostly to nationally funded research in Great Britain, however the research logic, as well as its instruments and methodology can be generalized and targeted to other domains and other countries, which we reserve to undertake in future work. In summary, the central problem this research tackles is that despite the existence of widespread open access policies which could appear *prima facie* to be in use, in the example of UK Research Councils, knowledge generated by Systems Engineering research using public funds is still not available to the public and sometimes not even to co-researchers.

## 4. Evidence of What Works

Knowledge reuse challenges can be examined under different disciplinary perspectives, but when tackled in combination, and considered 'as a whole', systemic traits such as 'entanglement' emerge.

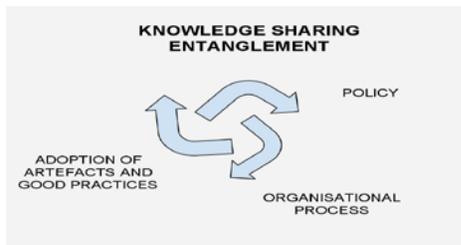

Image 1. Knowledge sharing entanglement

For example, knowledge sharing co-dependencies (entanglement) are addressed by relating two different dimensions of the problem space such as 'policy' and 'adoption of artifacts', constitutes the foundation of our mixed method research approach, as explained in related work (Di Maio, 2011) and illustrated schematically in image 1 above. The overall proposition that drives the inquiry is:

**There is a gap between the existence of the adoption of open access policies 'in theory' (T) and 'in practice' (P)**
from which the following questions and hypotheses are derived

**Q1. How can the gap between T and P be identified?**
**H1.** By gathering Evidence of the difference between T and P
**Q2. How can the gap between T and P be measured?**
**H2.** By devising indicators and parameters to evaluate the level of adoption
**Q3. How can the gap between T and P be reduced?**
**H3**. By devising and recommending appropriate measures and interventions

This paper provides an overview and a synthesis of findings obtained using different research methods, each contributing a piece of 'evidence' to help answer the question above, and to test the hypothesis. Evidence Based Research (EBR) emerges from a field known as 'Evidence based practice' (EBP):
*Evidence-based practice (EBP) method in the behavioral and social sciences developed out of the evidence-based movement in medicine, which aims to inculcate in clinicians "the conscientious, explicit, and judicious use of current best evidence in making decisions about the care of individual patients" (Sackett et al).*

The rationale for EBR is rooted in clinical practice in the health and medical domains, however a methodology has grown out of it, that has been adopted by other social science disciplines. It is noted (Paynter) that :
*While it may seem par for the course that professionals would use research to inform their practice, history is replete with examples of the opposite – practice based on the authority of their proponents rather than actual evidence of their efficacy. (Hatcher et al).*

A typical EBP research process consists of the following steps:
  (1) Formulate the question.
  (2) Search for answers.
  (3) Appraise the evidence.
  (4) Apply the results.
  (5) Assess the outcome
  (Gray, 2004)

This research, described in more detail in the sections that follow, complies with the central tenets of what constitutes a 'systemic review' method (EPPI) :
• Explicit and transparent methods are used following a standard set of stages.

• It is accountable, replicable and updatable.
• User involvement is built into the research design.

### 4.1 Research Design

The analytical part of study consists of two main research components, a Critical Appraisal of existing policies and legal instruments, and a systematic review of funded projects (the audits), that adhere to our inclusion criteria, described in more detail below.

1) A critical appraisal (evaluation) of Open Access and other Knowledge Sharing policies that guide or regulate academic practice in the UK (Davies), is aimed at answering the following questions

**Exploratory evaluation:** what policies are there? (Method: literature review, surveys and interviews with civil servants and experts)

**Impact evaluation**: do people know about these policies? (Method: audits, survey)

The steps undertaken in these research components are:

-   Identify public research funding bodies

- Survey and analyze their policy and implementation strategies

2) Survey/Audits of the Field. To look for evidence showing that the policies are implemented as intended, or otherwise (test the hypotheses) a survey of a targeted sample of the population and projects of UK academic projects is undertaken. A survey instrument is designed to carry out data collection for this sample called OAM (www.openaccessmonitor.org), further  specified in separate documentation linked on the site. The steps followed in this research component are:

- Select Cases to be audited (following the inclusion criteria)

- Gather data sets following OAM audit framework

- Analyse the findings according to multiple methods (qualitative, quantitative)

Inclusion criteria for the selection of cases in the current study are all the projects related to the target domain (in this study systems engineering research), UK-based and publicly funded through one or more UK research councils. The research concludes deriving from the findings a set of recommendations, which combine good practices with suggested interventions, such as policy integration and alignment, community involvement, and the adoption of suitable technical artifacts and knowledge models. It proposes a schematic reference model for shared knowledge representation, as well as other artefacts.

### 4.2 Motivation and Chain of Evidence

The initial motivation for the study was provided by NECTISE, a summary of the case is presented in the following section, as well as other observations, for which elements of ethnography were adopted. Academics (including principal investigators, researchers and postgraduate students) as well as practitioners showed little or no awareness of Open Access principles, which confirmed the findings of previous related studies (Swan). A series of interviews and email exchanges with experts followed to investigate various aspects of the problem space. The chain of evidence for this research is illustrated below:

NECTISE     >>> Initial observations
Ethnography >>>Awareness  of OA
Literature >>> Previous studies confirm observations
Survey, Interviews >>> Evidence from funding councils
Audits >>>  Systemic survey of the field (ongoing)

### 4.3 NECTISE

The underlying, endemic problem tackled by this research is well illustrated by one of the exploratory cases that initially triggered, and largely still motivates, most of this research: a portion of the EPSRC funded NECTISE (www.nectise.com), a 4 million GBP research project awarded to a consortium of prestigious Universities for 'networked capabilities in systems engineering', was allocated to investigate 'knowledge reuse'. As a doctoral training account holder (DTA) tasked with advancing the state of the art in 'knowledge reuse and learning in networked capabilities research for systems engineering' and receiving doctoral research funding from the NECTISE funding pool, it was essential for this researcher to acquire and examine existing project knowledge before the state of the art could be advanced further. However, the only knowledge resources publicly available via the NECTISE website were a static list of published papers (not hyper linked, nor available via the site, just enumerated on an HTML page on the project website).  Despite the fact that the project was publicly funded by EPSRC, due to contractual arrangements with industry partner BAE Systems. a private company which operates a policy of strict knowledge control, NECTISE never shared nor published any system diagrams, nor vocabularies or data dictionaries, and the research partners had to ask for permission to BAE before any decision could be taken. Although some of the papers linked on the project page could have been retrieved from scholarly repositories via web searches, they were mainly narratives and did not contain structured, systematic knowledge that could be re-used. An endless sequence of emails to obtain access to the knowledge artifacts related to the project between the doctoral researcher and entire research hierarchies of academics and officers in charge, generated no results - the 'target knowledge' was never obtained. No obvious open access project resources despite clear open access policies published by the funding body - prompted the question that motivates and justifies much of the current line of inquiry: if this research is publicly funded via EPSRC (Engineering and Physical Sciences Research Council), which like all other UK Research Councils embraces an 'open access' policy, why can't everyone, especially a researcher funded by the same project, access any of the project knowledge they need to carry our their professional or research task?

### 4.4 Knowledge Sharing Behaviours

An earlier pilot study, combined to international field work and an analysis of scholarly outputs, (Lee, Shiva) resulted in the identification of significant demographic differences that could contribute to shape the diversity of knowledge sharing behaviors For example a combination of factors, including Country, Job Role, Industry, Organisational Culture, can all impact to knowledge sharing attitudes and behaviors, a line of inquiry already partly explored in related research (Graf et al). While the theoretical part of this study is generalizable and domain independent (the research design and instruments can be modified to target different segments of the research field, or different research domains), given the constraint on resources it is necessary to narrow the current of scope of research to systems engineering research in the UK (see the gray cells in the table 1 below). The first research component, the critical appraisal of policy instruments, targets major research funding councils in the UK, considered in the context of related EU and international policies. The second part of the study, (the audits) has been targeted to systems engineering research projects in the UK

Table 1: Segmentation of the research field

|  | INDUSTRY | ROLE | SECTOR | POLICY |
|---|---|---|---|---|
| WORLD |  |  |  |  |
| EU |  |  |  |  |
| UK | systems engineering | researchers, research funding administrators | research | funding body, institution |

### 4.5 Open Access for Knowledge Sharing

The regulations, legislation and policies that govern 'knowledge sharing' practices in academia and industry are an entangled web of instruments, characterized by the tension between a global cyber-culture on the one hand, that promotes knowledge sharing and the adoption of web based artifacts to facilitate free and unrestricted knowledge flows, and on the other hand strong commercial interests of a 'knowledge economy' that can exist only via restrictions to knowledge via intellectual property rights enforced through commercial contracts, which enable the materialisation of earnings from Knowledge Transfer activities, such as for example the sale of books, course fees, etc. For the purpose of the analysis, the regulatory landscape has been segmented as follows:

- International declarations (OECD, Budapest, Berlin)
- International directives (EU PSI 2003)
- National legislation (that apply in a single member state to all governing bodies, such as the FOI Act 2000)
- National policies of each governing body
- Knowledge Transfer policies

A more detailed exploration of each of the segments above is being reported in a separate paper (under review, as of August 2011), however for the purpose of this paper, a brief summary of each segment is provided below.

### 4.5.1 International Declarations

Open Access is the broad term that identifies a progressive movement and a series of initiatives that have gradually lowered the barriers to access publicly funded research outputs. There is a long and rich history that documents how this movement evolved thanks to the efforts of individuals, groups and collectives that has finally been embraced at least to some extent by institutions (Suber). Key initiatives include the Budapest Open Access Initiative, Berlin and Bethesda, which yielded slightly different definitions, however the classic definition of reference is:

*"By "open access" we mean the free availability on the public internet, permitting any users to read, download, copy, distribute, print, search, or link to the full texts of these articles […] (Budapest Open Access Initiative, 2002)"*

Open Access to Scholarly Knowledge, has been a huge and growing movement, however it is noted that the initiatives above are not reflected in any legislation, and at the time of writing, no legislation exists that governs nor mandates the monitoring of open access publishing

### 4.5.2 International Directives

Public Sector Information has always been one of the main sources of primary data for many research activities and data centric services in modern economies, but thanks to the current explosion of web based technology applications and infrastructure that many more opportunities are opening up for a variety of stakeholders. The Council and the European Parliament adopted a Directive on the re-use of public sector information which deals with the way public sector bodies should enhance re-use of their information resources which, the EU says, is based on two key pillars of the internal market: transparency and fair competition (EU Council Directive of PSI Reuse 2003). The directive establishes minimum rules for the re-use of PSI throughout the European Union, but encourages Member States to go beyond minimum rules and to adopt open data policies, allowing a broad use of documents held by public sector bodies. Individual member states have adopted the directive with different legislative instruments and local variations (Implementation of the PSI directive). Interestingly, research institutions are excluded from the EU PSI directive with a comma in its Article 1

The Directive shall not apply to [...] *e) documents held by educational and research establishments, such as schools, universities, archives, libraries and research facilities including, where relevant, organisations established for the transfer of research results;*

Neither the EU PSI Directive of 2003 nor the UK Re-use of Public Sector Information Regulations 2005 specifically define what public sector information is. However, both the EU Directive and the UK Regulations make clear it covers information and content that is held by public sector bodies that fall within the scope of the Directive Regulations and where the rights are held by the public sector body. (private email exchanges with the press office of the Office of The National Archive, July 2011). For the purpose of FOI legislation however (at least in the UK) universities are considered 'public authorities' and must comply with FOI legislation. There seems to be a contradiction between the definition of public authority of the FOI Act and of the Regulations derived from the EU Directive.

### 4.5.3  National Legislation (UK)

The most notable example of legislation aimed at making accessible and transparent public sector information, is the FOI Act. In the UK publicly funded Research Institutions and Universities are considered 'public authorities', and therefore PSI legislation applies (FOI Act 2000)In the UK, the FOI Act and the Regulation 1515, both aimed at increasing 'access to knowledge' seem to be conflicting in their definition of public authority.

### 4.5.4  National policies of individual governing bodies (UK)

At national level, each governing body responsible for a public sector, may adopt a different version of the relevant policy. For example in the UK, each of the five major research councils have a different position in relation to a)open access b) data sharing. In further work, we reserve to undertake a more detailed comparative analysis of the same. Image 2 below provides a notional comparison of the policies, based purely on what the policies documents state. A closer evaluation, supported by the findings of our audits, reported in a later section of this paper, reveals that some of policy coverage stated 'on paper' cannot easily be verified: for example, EPSRC states on paper that it monitors the policy implementation, while according both to what emerged in the NECTISE case and to other audits, EPSRC at the time our investigation started, did not keep a record of Open Access resources for each funded projects. (Research Log Entry[1]). Further work is currently being undertaken to obtain evidence from corresponding organisations of monitoring activities, however, the criteria and extent for monitoring are unclear.

| Research Funders | Policy coverage | | Policy stipulations | | | | | Support provided | | | |
|---|---|---|---|---|---|---|---|---|---|---|---|
| | Published outputs | Data | Time limits | Data plan | Access / sharing | Long-term curation | Monitoring | Guidance | Repository | Data centre | Costs |
| Arts and Humanities Research Council | ● | ● | ● | ● | ● | ○ | ○ | ● | ○ | ● | ○ |
| Biotechnology & Biological Sciences Research Council | ● | ● | ● | ● | ● | ● | ● | ● | ● | ● | ● |
| Cancer Research UK | ● | ● | ● | ● | ● | ● | ● | ● | ● | ○ | ○ |
| Engineering and Physical Sciences Research Council | ● | ● | ● | ○ | ● | ● | ● | ● | ○ | ○ | ● |
| Economic and Social Research Council | ● | ● | ● | ● | ● | ● | ● | ● | ● | ● | ● |
| Medical Research Council | ● | ● | ● | ● | ● | ● | ○ | ● | ● | ○ | ● |
| Natural Environment Research Council | ● | ● | ● | ● | ● | ● | ● | ● | ● | ● | ● |
| Science and Technology Facilities Council | ● | ○ | ● | ○ | ● | ○ | ○ | ● | ● | ● | ○ |
| Wellcome Trust | ● | ● | ● | ● | ● | ● | ● | ● | ● | ● | ● |

Image 2: Curation Policies DCC Edinburgh[2]

### 4.5.5  Knowledge Transfer (KT)

Knowledge transfer (KT) can be used to describe the knowledge flows between different units, divisions, or organizations rather than individuals (Szulanski, Cappetta, Jensen), the emphasis of KT is on generating income from knowledge transfer activities, rather than maximising access to knowledge. KT is also defined as "the process through which one unit (e.g., group, department, or division) is affected by the experience of another" (Argote, Ingram). The EU Commission also states that it wants to move towards a position in which:

*"knowledge transfer between universities and industry is*

---

1  http://fieldnote.posterous.com/knowledge-reuse-in-systems-engineering
2  http://www.dcc.ac.uk/resources/policy-and-legal/overview-funders-data-policies

*made a permanent political and operational priority for all public research funding bodies within a Member State, at both national and regional level". (Commission Recommendation)*

As illustrated in Image 3, Knowledge Transfer principles consist of restricting access to knowledge, to allow for the commercial exploitation of knowledge resources, and generate income streams via the sale of educational materials (teaching), consulting services and licensing mechanisms - essentially in direct contrast with Open Access principles.

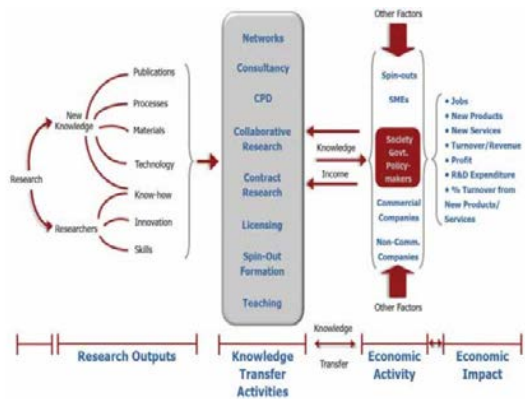

Image 3: Model of Knowledge Transfer within the Innovation Ecosystem (Source: University of Glasgow) In: Metrics for the Evaluation of Knowledge Transfer Activities at Universities (Unico Report)

Furthermore, the analysis of literature in the UK and EU reveals that Knowledge Transfer policies shape and mandate the knowledge exchange perceptions and behaviors at praxis level (Hauser). Intellectual Property clauses of commercial contracts part of 'Knowledge transfer programmes' restrict and constrain knowledge flows between academia and industry, effectively pre-empting Open Access policies to take hold. (Gardner, Fong, Huang). One of the asymmetries that become visible when contrasting of KT vs OA policies, is that the first are grounded in contract law, which is made firm in the law (contract law) while Open Access policies, at the time of writing, are still 'guidelines', so the first are prioritized due to their legal weight (Burnhill).The sections above provide an overview of the diverse set of initiatives, policies, and corresponding regulations that govern knowledge sharing practices, partly known as 'open access policies'. Findings and summary conclusions of this analysis of the landscape are presented in section 5 of this paper.

### 4.5.6 Auditing the Field

Despite the fragmentation of the regulatory landscape discussed above, each research funding council in the UK has their own 'open access policy' as reported in Image 2. The next step in our research process consisted of carrying out a systematic review of publicly funded projects, to see to what extent such policies were adhered to by principal investigators and their corresponding institutions. Although the digital curation community may not consider the distinction between information and knowledge, the so called 'knowledge level' (Newell) has different implications. An ad hoc Knowledge Auditing instrument was devised (KAF)[3] however this resulted to target 'too granular' level of knowledge - that is, KAF was devised using knowledge engineering principles aimed at specifying a high degree of formality of detailed technical knowledge. The KAF auditing process is illustrated in the image 4.

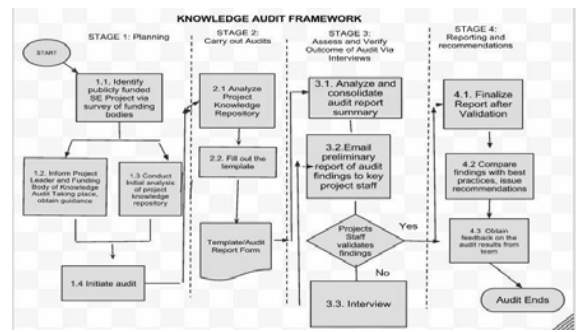

Image 4: Knowledge Audit Framework Process

After piloting KAF in the field, it emerged that the auditing instrument was over specified, for example it looked for systems specifications and diagrams, when it became clear that the majority of projects in the systems engineering research being audited did not even have a website, and of those which have a website, very few have links to accessible copies of deliverables and papers. Therefore a more generic, 'evidence based research' instrument evolved from KAF, called Open Access Monitor (OAM), a public version - slightly more polished instance of the data collection tool used in our audits for this research - is accessible on the web at http://www.openaccessmonitor.org. OAM consists of simple guiding principles, a process and data collection instruments (forms) and corresponding public data repositories to store the audited knowledge. It is designed to harvest a wider range of knowledge sharing standards, from the simplest form - 'does the project have a website'? - to more detailed, technical audit of knowledge sharing formalisms adopted - does the project open access resources have a unique web address (URI), and are they published using appropriate formalisms and notation?

---

[3] KAF http://tinyurl.com/3oleaaf

The current version of OAM is a working prototype developed at 'near zero cost', that is, using freely available development tools (Google apps). OAM evolved organically out of KAF, keeping the adopts its core process and inventorying mechanism, however, it uses an 'abbreviated protocol' (a simpler and less granular inventory process). OAM uses different inventorying templates to gather evidence about existing Open Access Policies (Policy Monitor) and about how publicly funded projects embrace the policies (Project Monitor). OAM also encourages and supports public intervention by providing an open, publicly accessible record of civic interventions (i.e. it logs email requests sent to funding bodies when Open Access resources in relation to a given project are not found). OAM provides a unified environment to assist knowledge auditors 'score' publicly funded research projects according a simple star schema which constitutes a form of 'heuristic evaluation' (Nielsen, J; Porter et al). The star rating systems is modeled on the linked data star system (Berners Lee). OAM 's internal architecture (the process and the templates) and methodology are available as documentation, however the star system is illustrated in the image below.

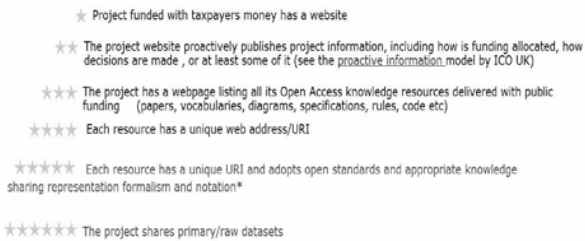

Image 5: Heuristic Assessment of KS via star system

Overall, OAM templates used in combination can help knowledge auditors answer the following questions:

1.  What are the Open Access policies of each funding body? How is the implementation of these policies monitored?

2.  Which Open Access knowledge resources are shared in the public domain for each publicly funded project? In addition, if a specific Open Access resource is not located, OAM encourages individual independent 'auditors' to write to the corresponding funding body, and to log such inquiry, related correspondence and responses in a public record.

### 4.5.7 The auditing sample

The target of the audits portion of the study are systems engineering research project in the UK, the funding council that specifically targets SE is the EPSRC, although other research councils such as the ATRC also funds large scale systems, they do not categorize 'systems engineering' as such. A comparative evaluation of different categorisation systems for different research councils points to the need to further harmonize, or at least map, the conceptual and categorization schemas for different councils, but we leave this discussion for a future work. Given the relatively contained number - approximately 100 - of systems engineering research projects funded by EPSRC that ended in 2010 and 2009 it was decided the sampling strategy was a 'census', ie, it did not require a selection of a subset of the total sample, but given existing resources, and by recruiting volunteer auditors, they could all be audited. It should be noted that since the audits took place while OAM was still in development, only five of the six criteria were audited in our study (the sixth criterion was added later).

## 5. Results

Below a summary of preliminary findings to date, corresponding to each research components: policy evaluation (theory) evidence from the field (practice)

### 5.1 Policy Evaluation Findings

The policy assessment effort was initiated as part of this research with the goal of understanding what OA policies exist, and to what extent funding councils implement and monitor them. Different methods for policy evaluation were adopted in combination (Purdon et al). Outcomes of this evaluation point to the following conclusions:

1. The policy landscape is fragmented across different levels. For example, different policies addressed loosely different layers of the information management chain, for example: Data, Information, Knowledge.

2. There are different policies with different scopes and purposes, all targeting roughly the same 'knowledge sharing' space, but which are not harmonized,

3. Some of the current legal provisions for the protection of Intellectual Property, and programmes such as 'Knowledge transfer' that restrict knowledge flows between academia and industry, could be in conflict with Open Access policies.

4. UK Funding bodies have Open Access policies in place, however they do not monitor, and when they do, they do not specify 'how' they monitor the implementation of OA policies.

## 5.2 Field Research Findings

The first observation, via the NECTISE case, provided the initial evidence that motivated the rest of the study. Pilot interviews were undertaken, which resulted in one of the important 'lessons learned': structured surveys and questionnaires may not result in honest, truthful replies. Respondents are intimidated by the technical jargon in use (what is a formalism? was one of the typical reactions) and were also reluctant to share openly their attitudes and behaviors. An ethnographic approach was therefore adopted from that point onward, and systems engineers were observed in the course of the three year in various occasions via

1. participation in UK and international workgroups such as INCOSE
2. participation in international systems engineering conferences and events (UK and international)
3. direct observation and participation in international systems engineering projects (Incose SEBOK)

The aim of ethnographic observation is to gain some understanding of:
- do systems engineering researchers know what is open access?
- do they know what the Budapest Initiative is?
- do they know what knowledge sharing policies govern their publicly funded research?

One of the ethnographic studies consisted of casual 'on the spot' interviews carried out in 2009, where academics (researchers and postgraduate students) were asked in their natural work environment, and in the context of routine 'reseach interest' type of conversations, whether they knew what is Open Access, and what is the Budapest Declaration; One of the studies was carried out on campus (an engineering faculty in the UK). Of 30 participants, selected randomly (were physically approached on campus when the opportunity arose) and anonymously (their names were not recorded) all answers were negative: nobody knew what Open Access is, nor what the Budapest Declaration is. The same ethnographic experiment was repeated across a variety or events, over a period of time, with slight variations in the results.

Table 2 summarises the type of events and dates, number of subjects who were approached and their responses to the three questions above.

Although of limited statistical significance, these result point clearly to lack of awareness of open access. The findings on our limited sample confirmed the outcomes of earlier reports (Swan).

Table 2: Summary of ethnographic study

| SETTING | DATE | Nr | Q1y | Q1n | Q2y | Q2n | Q3y | Q3n |
|---|---|---|---|---|---|---|---|---|
| walk in engineering campus (6 weeks, local) | 2009 | 30 | 0 | 30 | 0 | 30 | 0 | 30 |
| systems engineering networking meeting (1 day, national) | 2010 | 24 | 0 | 24 | 0 | 24 | 0 | 24 |
| Space symposium (local) | 2010 | 14 | 0 | 14 | 0 | 14 | 0 | 14 |
| Syseng Iternational conferences (4 days, international) | 2010 | 30 | 4 | 26 | 0 | 30 | 5 | 25 |
| Syseng National Conference (2 days, international) | 2011 | 22 | 3 | 19 | 2 | 20 | 4 | 18 |

It was therefore decided that no further data was needed to demonstrate the 'lack of awareness' problem. Instead, a systematic survey of publicly funded projects in the SE domain was undertaken using OAM. Four initial pilot audits were carried out, which helped refine the monitoring instrument and fine tune the auditing procedures. A total of 100 EPSRC funded projects ended in 2009 and 2010 has been audited and 'scored' to date, with the following results: the majority of project audited did not have open access knowledge resources, or very few (<3), however the good news was that the third largest group of 11 audits scored very high (>14).

Table 3: OAM Scores

| Number of Audits | Score |
|---|---|
| 57 | 0 |
| 15 | 1 |
| 1 | 2 |
| 5 | 3 |
| 5 | 6 |
| 6 | 10 |
| 2 | 14 |
| 11 | 15 |

The pie chart below represents diagrammatically the figures in the table

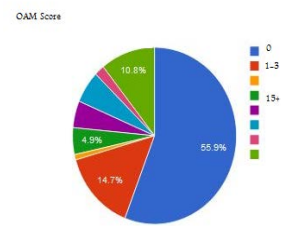

Additional datasets with some variance in the inclusion criteria are being gathered to permit further analysis, however from the current findings, the following conclusions can be drawn:

1) The majority of projects audited did not have any or very few open access resources,

2) Almost all projects have some papers published that can be retrieved via web searches associated to the grant number

3) The third largest segment of the audited population (approx 10%) adheres to all good practices and knowledge sharing conventions (>14)

4) The quality, detail and sharing formalisms adopted by each project varies greatly and does not always depend on the existence of an explicit knowledge sharing policy. Web page, when they exist, are used mainly as promotion/marketing, and no consistent KS formalization is adopted.

5) Other factors, such as background of the grant holder and organisational culture may contribute to the level of granularity and knowledge sharing formalisms adopted.

6) the quality and type of knowledge shared in publicly funded systems engineering research tends to be high level, narratives (papers), however limited formalized and reusable system knowledge is routinely published and shared.

7) The minority of projects audited that adopt standard knowledge sharing practices (the notable exceptions) do so consistently and in compliance with good practices. Currently each of these projects is being used as a 'model of good practice', and studied more closely, to gather additional insights into outstanding KS behavior. In summary, the evidence gathered so far from field work points toward the following conclusions:

- Researchers in systems engineering are generally not aware of OA policies.

- Only a limited number of publicly funded projects complies with the policies of their funding bodies

- Open Access policies are underspecified and vague.

A number of other qualitative considerations that have emerged from the evaluation of the findings as a whole are currently being elaborated in a technical report that will be sent to all individuals and institutional representatives, and that will serve as the basis for further research.

## 6.    Recommendations

Over the course of the study, initial evidence and findings were discussed and presented to various individuals in selected funding bodies and organizations, some of which are logged as research notes (research log, private correspondence). During the course of these exchanges, a press statement was issued by Research Councils UK (RCUK) and the Higher Education Funding Council for England (HEFCE)[4] announcing plans to work together to ensure greater open access to published research (Announcement, 25 May 2011). In particular, the EPSRC, the research council with which we have had intense correspondence exchanges for the last two years, circulated a policy update  (Ryan, B) which states textually:

*'EPSRC will monitor compliance with the policy as part of our normal assurance processes'*

However there is no indication as to what the open access monitoring strategy going to be, and more importantly, no indication of what level of public resources are going to be devoted to this effort, in simple terms, it is not clear how much the monitoring of open access resources is going to cost the tax payer, nor how efficient and effective is going to be. One of the contributions of this research is a set of recommendations gathered partly from standard good practices, as we learn them from web science and knowledge engineering, and partly emerge from the empirical evaluation of the evidence gathered in the course of the study. These recommendations are grouped into four  distinct categories:
- toward a reference model for knowledge sharing
- for governing councils, research funding bodies
- for institutions
- for individual researchers

'Open access' is a broad, boundary spanning complex socio-technical challenge, and the proposed recommendations are best adopted in combination: simple, cost effective measures articulated across the different levels of the problem space can yield systemic results.

## 6.1    Towards a General Reference Model for Knowledge Sharing

To achieve optimal knowledge sharing potential of codified knowledge resources, such as technical knowledge, it is necessary to adopt appropriate conventions, formalisms and artifacts. Some of these conventions are well established, and  have been encoded as a knowledge sharing star rating system for  OAM  . however no single knowledge schema exist that researchers can adopt when trying to make their outputs more useful, and more easily accessible.  The rationale and workplan toward the development of a reference ontology and a shared vocabulary for the system engineering practice, is reported in a separate paper, (Di Maio,  Proceedings of the ACM, 2011). The outcome of the knowledge and content analysis of knowledge resources in the systems engineering domain has resulted in a sample 'reference model' reported below, whereby the system development phases correspond knowledge artifacts, logically articulated, represented and shared using appropriate formalisms, notation and file formats. Similar domain dependent knowledge reference schemas can be developed and adopted in other fields of practice.

---

[4] http://www.rcuk.ac.uk/media/news/2011news/Pages/110525_1.aspx

Reference Model of Knowledge Sharing in SE (Di Maio)

| LIFECYCLE PHASE | KNOWLEDGE ASSET *document, specification | FORMAT | NOTATION/FORMALISM | SHARING MECHANISM |
|---|---|---|---|---|
| analysis | requirements specification | narrative structured text | natural language, pseudo-code | image word document spreadsheet pdf html xml rdf owl other |
| design | system diagram | diagram | ER, DF, UML | |
| development | system specification | narrative structured text | Natural language pseudocode | |
| installation | operating manual user guide | narrative diagrams | Natural language graphics | |
| testing | test plan | structured text | natural language charts | |
| acceptance | contract | narrative | natural language | |
| support | user feedback tickets feedback | narrative | natural language | |

The overall general recommendation - as well as one of the contributions of this research to the systems engineering research domain - is that basic traditional web knowledge sharing artifacts and core practices, such as the use of URIs for sharing knowledge resources, appropriately used address and resolve most knowledge sharing challenges. Furthermore, adopting a domain specific knowledge reference model, as illustrated in the table above, can mitigate at least in part the lack of more sophisticated shared codification standards.

## 6.2   Fact Checking

'Scientific knowledge' rests, above all, on facts, whereby science itself is about verifiability and reproducibility. This research is developed in the context of an engineering discipline, in particular systems, web and knowledge engineering, whereby engineering is intended as 'the practical application of science to commerce or industry' [Fox]. The ability to verify facts via gather evidence is essential to reason, make inferences, draw conclusions and essentially, to make informed decisions. On the web, which is the largest open, large scale distributed knowledge base, fact checking is particularly important to the accuracy of reasoning, which can be defined as the act or process of using one's reason to derive one statement or assertion (the conclusion) from a prior group of statements or assertions (the premises) by means of a given method [Clarke]. The validity of 'knowledge' requires it to be verified or verifiable, with some exceptions that may be satisfied with theoretical assumptions. Fact checking is adopted routinely in investigations (research) in providing evidence (legal/making the case) and in decision making (to reduce over reliance on assumptions). It is recommended that when sharing knowledge on the web, the mechanism to provide verifiable evidence is to use hyper links to corresponding documents, which can be either HTML or RDF. In related work, the linked data model is explored as a possible formalization for fact checking [5]

## 6.3   For policy makers and funding bodies

The fragmented state of heterogeneous policies and legislation can be confusing, and even lead to contradictory practices, as identified in the relevant section of this paper. Although it is acceptable to have multiple policies, it would be advisable a certain level of cohesion, integration and alignment between them.

a) An open access policy management strategy should enable dual track, i.e. encourage compliance from the bottom up (self archiving) but also encourage funding bodies and regulators to implement the policy via regulatory measures (mandates) and above all monitor compliance with the policy

b) bridge the current fragmentation between data, information and knowledge policies, and establish a firm 'correspondence' between the policy and the mandates on the one hand, which can be called the social and organisational aspects of knowledge sharing, and the adoption of the knowledge sharing artefacts, conventions and standards, that can be defined as the technical aspects, because the two are facets of the 'same coin', as shown diagrammatically in the illustration below.

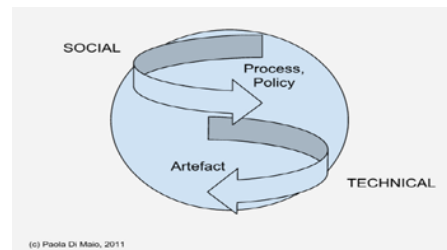

Image 6: Socio Technical Approach to Open Access

c) devise and implement an overall integrated Open Access policy monitoring strategy which should be in line, and where possible extend, the guidelines provided by international directives, such as EU PSI Directive 2003.

d) Leverage the community: it is expected that budgetary considerations will play a role in how effectively the monitoring of Open Access policies implementations will be. If carried out manually, and without use of ICT, the cost of monitoring policy implementation could exceed its benefits. However, if a simple automated policy monitoring process is implemented by mandate, say via an open web service such as OAM, the burden of monitoring could be distributed across the research community or even crowd-sourced which would reduce the material costs of a much needed monitoring to almost zero.

---

5   Provenance and Linked Data Workshop, SICSA, University of Edinburgh 2011

d) issue clear guidelines as to what level of data, information and knowledge should be made freely accessible to by mandate, and which levels can be protected by patents and copyright to allow research outputs to benefit from commercialization opportunities and economic gain via Knowledge Transfer agreements

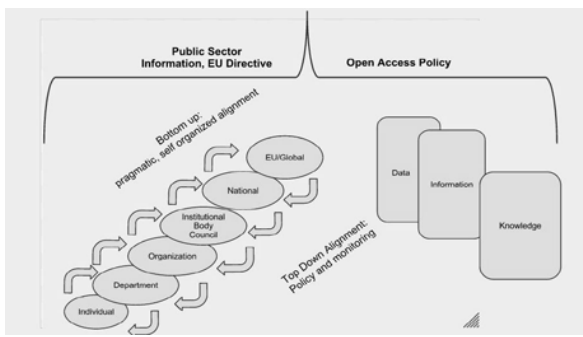

Image 7: Intervention: Integration and alignment of the fragmented policies regulating the space

d) consider legislation. At the moment, the provisions for commercial knowledge transfer are entered contractually, whereby contracts are legally binding instruments entered enforced by contract law. Open access policies are still operated as guidelines only, and carry no legal, binding weight. The relation between open access policies and knowledge transfer agreements is strongly asymmetrical in the law, and favors the latter.

## 6.4  For research institutions

Institutions, as large bureaucratic organisations, tend to be 'passive', and to follow directions issued from 'the top' by governing bodies. When a policy carrying strategic implications for the advancement of science at global and national level, such as the policy for open access to scholarly publications, it is necessary for everyone in the research supply chain to wholeheartedly embrace it. What good is a policy emitted by a funding body, if no institution adheres to it? Institutions have primary responsibilities toward the public at large, as well as toward the public funding councils, and the research community. Their responsibility is to understand the open access framework, and to pass it on to their entourage. The primary recommendations for research institutions are as follows:

a) embrace the culture of knowledge sharing. this often implies a disruptive overhaul of pre-constituted knowledge hierarchies, and it cannot be achieved overnight

b) provide regular training about knowledge sharing and where necessary technical support for researchers

c) issue guidelines and recommendations as to what optimal knowledge sharing practices are, including recommending the adoption of existing artifacts and good practices, and stimulate the innovative development of new ones.

## 6.5  For Researchers

In contemporary networked society, self governance, as well as the active participation of individuals in all governance practices of institutions, is encouraged, but this cannot happen without the researchers understanding the political and practical implications of information policies.

1. Publish often, as often as possible, and do not wait for results to be complete and exhaustive, share your findings early, update the findings with progress reports.

2. Share your knowledge and data using standard good practices, contribute to the development of the same.

3. Favor, where possible, working for institutions transparent and compliant with good practices and support open access

4. Contribute to the active evangelization and monitoring of open access in your research environment, and become a point of reference for your community.

## 7.  Contribution, and Future Work

This research so far claims the following contributions:
- the first systematic review of open access in the systems engineering research
- the first 'evidence based research' contributed to systems engineering research
- contributes the novel concept and example of 'heuristics evaluation' to knowledge sharing research (the star system)

Additional data cross validation for ancillary quantitative analysis of the findings is currently being undertaken. Future work includes a wider study using OAM in other domains and countries, a contribution to public consultations both in the UK and the EU.

## 8. Conclusion

This paper presents the rationale, methodology and some of the findings and recommendations of a study aimed at filling the gap between Open Access theory and practice. It introduces OAM, a near zero cost public environment to support the monitoring of open access policies and presents an example of 'reference model' for knowledge sharing in systems engineering.

## 9. Acknowledgments

This research is partly funded with EPSRC Grant nr. EP/D505461/1 for the project. "Network Enabled Capability Through. Innovative Systems Engineering (NECTISE). Gratitude to many people who are providing essential feedback and guidance in handling the complex intricacies and dependencies of the diverse components that make this research possible

Paola Di Maio holds a BA Hons (1994) and an MSc (2000), is a research analyst for Cutter.com, independent expert for the European Research Agency (REA). She works as a standards evaluator and research advisor, is a Research Associate at Institute of Socio-technical Complex Systems in the UK, lectures internationally.